\documentclass[pre,showpacs,floatfix,twocolumn]{revtex4-1}

\usepackage{graphicx}
\usepackage{subfigure}
\usepackage{amsmath}
\usepackage{amssymb}
\usepackage{latexsym}
\usepackage{multirow}

\newcommand{\ve}[1][K]{\mathbf{#1}}

\begin{document}

\title{Force-induced dispersion in heterogeneous media}

\author{T. Gu\'erin}
\author{D. S. Dean}
\affiliation{Laboratoire Ondes et
Mati\`ere d'Aquitaine (LOMA), CNRS, UMR 5798 / Universit\'e de  Bordeaux, F-33400 Talence, France}

\date{\today}

\begin{abstract}
The effect of a constant applied external force, induced for instance by an electric or gravitational field, on the dispersion of Brownian particles in periodic media with spatially varying diffusivity, and thus mobility,  is studied. We show that external forces can greatly enhance dispersion in the direction of the applied force and also modify, to a lesser extent and in some cases non-monotonically, dispersion perpendicular to the applied force. Our results thus open up the intriguing possibility of modulating the dispersive properties of heterogeneous media by using externally applied force fields. These results are obtained via a Kubo formula which can be applied to any periodic advection diffusion system in any spatial dimension.
\end{abstract}


\maketitle
In diverse systems ranging from  fluid mechanics, hydrology, soft matter to solid state physics, at mesoscopic length and time scales, the dynamics of tracer particles is described by stochastic differential equations (SDEs) and their associated Fokker-Planck equations \cite{VanKampen1992,oksendal2003stochastic,gardiner1983handbook}. In heterogeneous media, the local transport coefficients such as the diffusivity and the mobility can vary in space depending on the local material properties. In a locally isotropic material  where a uniform force ${\bf F}$ acts on a tracer particle, the probability density function (PDF) $p({\bf x},t)$ for the tracer position at time $t$ obeys 
\begin{align}
	\partial_t p({\bf x},t)= \nabla\cdot\left[\kappa({\bf x})\nabla p - \beta\ \kappa({\bf x})\  {\bf F}\ p\ \right]. \label{eqkF}
\end{align}
The first term on the right hand side of Eq.~(\ref{eqkF})  above corresponds to diffusion with a spatially varying diffusion constant. The second term represents the drift due to a constant applied external force and the term $\beta\kappa({\bf x})
=\mu({\bf x})$ is the local mobility. The factor of the inverse temperature $\beta$ results from the local Einstein relation 
between mobility and diffusivity. Physical examples include charge carriers in heterogeneous media, where $\mu({\bf x})$ is proportional  to the local electrical conductivity,  in the presence of an external electric field, as well as  colloidal diffusion in porous media, with local diffusivity $\kappa({\bf x})$, with an external field induced by gravitational or buoyancy forces. Here we study the effect that a constant external applied field has on the late time dispersion as characterized by the effective drift of a cloud of tracer particles
\begin{equation}
V_{i} = \lim_{t\to\infty} {\langle X_i(t)-X_i(0)\rangle\over t}, \label{DefVi}
\end{equation}
(where ${\bf X}(t)$ denotes the position of a tracer particle and $\langle \cdot\rangle$ denotes ensemble averaging) and the effective diffusivity
\begin{equation}
D_{ii}=\lim_{t\to\infty} {\langle[ X_i(t)-X_i(0)]^2\rangle_c\over 2t},\label{DefDii}
\end{equation}
($c$ denotes the connected part, thus the variance of the displacement $X_i(t)-X_i(0)$) characterizing the dispersion the cloud about its mean position.  Effective transport coefficient are important  for estimating the spread of pollutants and chemical reaction times \cite{condamin2007}.

When ${\bf F}={\bf 0}$, the problem of determining  $D_{ii}$ and $V _i$ dates back to Maxwell \cite{MaxwellBook}, where the equivalent problem of determining the dielectric constant of heterogeneous media was addressed. The Wiener bounds \cite{wiener1910} state that $(\overline{\kappa^{-1}})^{-1}\leq D\leq \overline\kappa$, where $\overline{\ \cdot\ }$ indicates spatial averaging. In higher dimensions there are few exact results \cite{dykhne1971conductivity} but numerous approximations schemes exist \cite{jeffrey1973,drummond1987effective,deWit1995correlation,abramovich1995effective,dean2008self}. However, the case where there is a finite external force appears not to have been studied  and in this Letter we will address the force's effect on the dispersion of tracer particles.

\begin{figure}[h!]
 \includegraphics[width=8cm,clip]{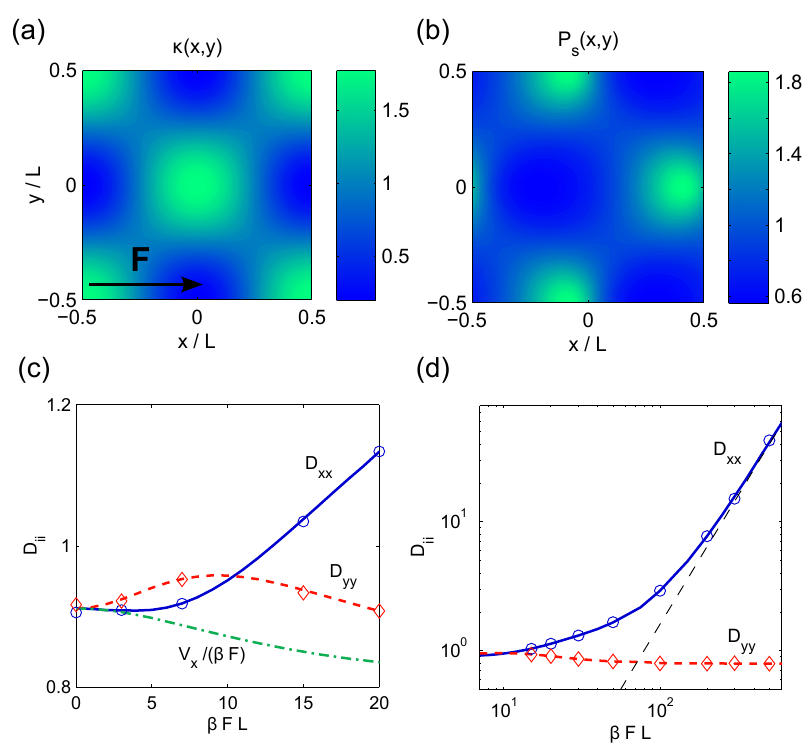}   
 \caption{(color online)  (a) The 2D periodic diffusivity field $\kappa(x,y)=\kappa_0[1+0.8\cos(2\pi x/L)\cos(2\pi y/L)]$, in units of $\kappa_0$ on the fundamental rectangular unit cell. The arrow indicates the direction of the external force. (b) Stationary PDF  in the diffusivity field shown in (a) with an external force of magnitude $\beta FL=10$. (c) Components $D_{xx}$ and $D_{yy}$ of the effective diffusion tensor predicted by Eqs.~(\ref{FunctionF},\ref{DispersionFunctionOfF})  and the normalized effective drift $V_x/\beta F$ from Eq.~(\ref{strat}) (lines) along with simulations results  for the SDE  (\ref{SDE_Ito}) (symbols). (d) Same as (c) with different scales. The dashed line represents the behavior $D_{xx}\simeq c F^2$ with the coefficient $c$ predicted by Eq.~(\ref{DxxForceSquared}).}
\label{FIG1}
\end{figure}  

To gain a flavor for the phenomenology of this problem we consider diffusion in a two
dimensional medium, where $\kappa(x,y)$ is shown in Fig.~\ref{FIG1}(a), with an applied force ${\bf F}$ oriented in the $x$ direction. We show in Fig.~\ref{FIG1}(c) the results of  numerical simulations of the corresponding SDE for the quantities $D_{xx},\ D_{yy}$ and $V_x/\beta F$. At zero force,  all the quantities shown are equal, this is a result of the Stokes-Einstein relation  $D_{xx}={\beta \partial_F V_x} $ which holds {\em only} \cite{commentStokes} when $F=0$. At small $F$ upon increasing $F$, we see that both $D_{xx}$ and $V_x/F$ decrease while $D_{yy}$ increases. As $F$ increases further, $V_x/F$ continues to decrease monotonically, however $D_{xx}$ and $D_{yy}$ attain minimal and maximal values respectively and eventually cross. This remarkable behavior shows that the fast and slow directions of dispersion can be interchanged by an applied force and that $D_{yy}(F)$ is a non-monotonic function. In Fig.~\ref{FIG1}(d), we see that $D_{xx}$ grows as $F^2$ at large forces and can thus be made arbitrarily large (thus exceeding the upper Wiener bound for the forceless case), giving rise to force induced dispersion enhancement. The key difference between systems with and without an external force is that in the latter case the steady state probability distribution $P_s(x,y)$ on  the periodic unit cell of the system is constant, whereas in the presence of the field it becomes non-trivial as shown in Fig.~\ref{FIG1}(b). 

To explain these results we will derive a  Kubo-type formula for the transport coefficients for general Fokker-Planck equations with arbitrary periodic diffusion tensors and advection fields. This formula generalizes a number of existing results for convection by incompressible velocity fields with constant molecular diffusivity as in the case  of Taylor dispersion \cite{taylor1953dispersion}. Examples include diffusion in Rayleigh-B\'enard convection cells \cite{rosenbluth1987effective,Shraiman1987,McCarty1988}, diffusion in frozen turbulent flows \cite{majda1999simplified} and  transport by a fluid in porous media \cite{brenner1980dispersion,souto1997dispersion,alshare2010modeling,carbonell1983dispersion}. Our formula also encapsulates results  for diffusion in periodic potentials \cite{dean2007effective,derrida1983velocity,zwanzig1988diffusion,deGennes1975brownian,lifson1962self}. In one dimension, results on diffusion in periodic potentials plus constant forces have been derived \cite{reimann2001giant,reimann2002diffusion,reimann2008weak,Machura2005,lindner2001optimal}, as well as  the more general case where the noise amplitude is a periodic function of  position \cite{lindner2002,reguera2006entropic,burada2008entropic}. 

The Kubo formula we derive here is valid in any dimension. The terms  in the Kubo formula can be analytically evaluated when the diffusivity varies only in one direction, and we give analytical results for such stratified systems. We also solve the generic problem analytically in the limit of large forces, proving that the coefficient of  $D_{ii}$,  where $i$ is the direction of the force, is  generically proportional to $F^2$. Finally, the Kubo formula can be evaluated  by solving a set of associated partial differential equations numerically \cite{commentStokes}, the excellent agreement between  this calculation and the simulations is shown in Figs.~\ref{FIG1}(c,d).

\textit{Kubo formula for the dispersion. } 
Consider the general Fokker-Planck equation 
\begin{align}
	\partial_t p =\sum_{i,j=1}^d\partial_{x_i}\{ -u_i(\ve[x])p+\partial_{x_j}[\kappa_{ij}(\ve[x])p]\}\equiv \mathcal{L}_{\ve[x]}\ p,\label{FKPEq}
\end{align}
where $\kappa_{ij}(\ve[x])$ is a local (symmetric) diffusion tensor, $\ve[u](\ve[x])$ is the  drift field, and $\mathcal{L}_{\ve[x]}$ is the transport operator. Our only assumption in the following is that the fields $u_i(\ve[x])$ and $\kappa_{ij}(\ve[x])$ are periodic in space. Let $\Omega$ denote the fundamental unit cell of the periodic structure. We call $p(\ve[x],t\vert \ve[y])$ the propagator of the stochastic process in infinite space, defined as the solution of Eq.~(\ref{FKPEq}) in infinite space with initial condition $p(\ve[x],0\vert \ve[y])=\delta(\ve[x]-\ve[y])$. We distinguish this infinite space propagator  $p(\ve[x],t\vert \ve[y])$ from the propagator  calculated with periodic boundary conditions on the boundaries of $\Omega$, denoted $P(\ve[x],t\vert \ve[y])$, and representing the probability density to observe a particle at time $t$ at a position $\ve[x]$ modulo an integer number of translations along the lattice vectors of the periodic structure. Finally, we define $P_{\mathrm{s}}(\ve[x])=\lim_{t\rightarrow\infty}P(\ve[x],t,\vert \ve[y])$ the stationary PDF of the particles with periodic boundary conditions. 

In the Ito prescription, the SDE corresponding to the Fokker-Planck equation~(\ref{FKPEq}) in the direction $i$ \cite{gardiner1983handbook,oksendal2003stochastic} is
\begin{align}
	dX_i=u_i(\ve[X](t)) \ dt + \sum_{j=1}^d(\kappa^{1/2}(\ve[X](t)))_{ij}dW_j \label{SDE_Ito},
\end{align}
where $\kappa^{1/2}$ represents the square-root matrix of the positive symmetric matrix $\kappa$.  
The  noise increments $dW_i$ are Gaussian, independent, of  zero mean and are only correlated at equal times as  $\langle dW_i dW_j\rangle=2\delta_{ij}dt$.
Ensemble averaging Eq.~(\ref{SDE_Ito}) yields the Stratonovich result    \cite{stratonovich1958oscillator}
\begin{align}
V_i  =   \int_{\Omega} d\ve[x] \ P_{\mathrm{s}}(\ve[x])\  u_i(\ve[x]). \label{strat}
\end{align}
To calculate the  effective diffusivity we first substract $u_i dt$ from both sides of Eq.~(\ref{SDE_Ito}), integrate over time,  square both sides of the resulting equation and then  average to find
\begin{align}
	\langle [X_i(t)-X_i(0)]^2\rangle&+  \int_0^t dt_1 \int_0^t dt_2   \langle u_i(\ve[X](t_1))u_i(\ve[X](t_2)) \rangle \nonumber\\
-2 \int_0^t dt' \langle \{X_i(t)&-X_i(t')+X_i(t')-X_i(0)\}u_i(\ve[X](t'))\rangle\nonumber\\
& = 
 2 t\int_{\Omega} d\ve[x] \ P_{\mathrm{s}}(\ve[x])  \kappa_{ii}(\ve[x]).\label{ExpansionSquareDx}
\end{align}
The average of the right hand side of Eq.~(\ref{ExpansionSquareDx}) follows from the independence of the  $dW_i$ at different time steps. Exploiting the periodicity of the field $\ve[u](\ve[x])$, we can evaluate the second term of Eq.~(\ref{ExpansionSquareDx}) for $t_1<t_2$ as
\begin{align}
 &  \langle u_i(\ve[X](t_1))u_i(\ve[X](t_2)) \rangle =\nonumber \\
&  \iint_{\Omega} d\ve[x]_1  d\ve[x]_2 u_i(\ve[x]_2)u_i(\ve[x]_1) P(\ve[x]_2,t_2-t_1\vert \ve[x]_1)P_{\text{s}}(\ve[x]_1) .
\end{align}
The second line of Eq.~(\ref{ExpansionSquareDx}) contains the term \cite{commentkubo}
\begin{align}
\langle [X_i&(\tau)-X_i(0)]u_i(\ve[X](0))\rangle = \nonumber\\
& \int_{\mathbb{R}^d} d\ve[x]\int_{\Omega} d\ve[y] \ p(\ve[x],\tau\vert \ve[y]) P_{\mathrm{s}}(\ve[y]) (x_i-y_i) u_i(\ve[y]). \label{ExpressionCorruu}
\end{align}
Differentiating with respect to $\tau$, using Eq.~(\ref{FKPEq}) and  integrating by parts over $\ve[x]$, we obtain 
\begin{align}
&\partial_{\tau}  \langle [X_i(\tau)-X_i(0)]u_i(\ve[X](0))\rangle =\int_{\Omega} d\ve[y] \ P_{\mathrm{s}}(\ve[y]) \  u_i(\ve[y]) \times \nonumber\\
&   \int_{\mathbb{R}^d} d\ve[x]\Big[u_i(\ve[x]) p(\ve[x],\tau\vert \ve[y]) - \sum_{j=1}^d\partial_{x_j}\kappa_{ij}(\ve[x])p(\ve[x],\tau\vert \ve[y])\Big]. \label{9457150}
\end{align}
Finally, exploiting the periodicity of the field $\ve[u]$, we can replace the integral over $\ve[x]$ over the infinite space by an integral over the unit cell $\Omega$ if one replaces the infinite space propagator $p$ by the propagator with periodic boundary conditions $P$, yielding for any $t>t'$ \cite{commentkubo2}
\begin{align}
\partial_t \langle &[X_i(t)-X_i(t')]u_i(\ve[X](t'))\rangle =  \nonumber\\
&\int_{\Omega} d\ve[x]\int_{\Omega} d\ve[y]   \  u_i(\ve[y]) u_i(\ve[x]) P(\ve[x],t-t'\vert \ve[y]) P_{\mathrm{s}}(\ve[y]). \label{958I2}
\end{align}
The last term to be computed in Eq.~(\ref{ExpansionSquareDx}) is 
\begin{align}
\langle [X_i&(t)-X_i(0)]u_i(\ve[X](t))\rangle = \nonumber\\
& \int_{\mathbb{R}^d} d\ve[x]\int_{\Omega} d\ve[y] \ p(\ve[x],t\vert \ve[y])P_{\mathrm{s}}(\ve[y]) (x_i-y_i) u_i(\ve[x]).
\end{align}
Due to the periodicity, we can exchange the integration domains of $\ve[y]$ and $\ve[x]$ in this equation. We now use the backward Fokker-Planck equation  \cite{gardiner1983handbook}  $\partial_t p(\ve[x],t\vert \ve[y])=\mathcal{L}_{\ve[y]}^{\dagger}p$, (where $\mathcal{L}^{\dagger}$ is the adjoint of the transport operator $\mathcal{L}$) to find
\begin{align}
 &\partial_{t} \langle [X_i(t)-X_i(0)]u_i(\ve[X](t))\rangle =  \nonumber\\
&\int_{\Omega} d\ve[x]  \int_{\mathbb{R}^d} d\ve[y] \ [\mathcal{L}_{\ve[y]}^{\dagger}p(\ve[x],t\vert \ve[y])] P_{\mathrm{s}}(\ve[y]) (x_i-y_i) u_i(\ve[x]).
\end{align}
Using the definition of the adjoint operator, we write
\begin{align}
& \partial_{t} \langle [X_i(t)-X_i(0)]u_i(\ve[X](t))\rangle = \nonumber\\
& \int_{\mathbb{R}^d} d\ve[y]\int_{\Omega} d\ve[x] \ u_i(\ve[x]) p(\ve[x],t\vert \ve[y]) \mathcal{L}_{\ve[y]}  \{P_{\mathrm{s}}(\ve[y]) (x_i-y_i)\} .\label{9582391}
\end{align}
Again exploiting the  periodicity of $\ve[u]$ and explicitly calculating  $\mathcal{L}_{\ve[y]}  \{P_{\mathrm{s}}(\ve[y]) (x_i-y_i)\}$ gives
\begin{align}
& \partial_{t} \langle [X_i(t)-X_i(0)]u_i(\ve[X](t))\rangle = \int_{\Omega}d\ve[x]\  u_i(\ve[x])\times \nonumber\\
 &\int_{\Omega} d\ve[y]   P(\ve[x],t\vert \ve[y]) \Bigg\{J_{\mathrm{s},i}(\ve[y])-\sum_{j=1}^d\partial_{y_j} [ \kappa_{ij}(\ve[y])  P_{\mathrm{s}}(\ve[y])]\Bigg\} ,\label{049141}
\end{align}
where $\ve[J]_{\mathrm{s}}(\ve[y])$ the local current in the stationary state at position $\ve[y]$, given by
\begin{align}
J_{\mathrm{s},i}(\ve[y])= u_i(\ve[y]) P_{\mathrm{s}}(\ve[y]) -\sum_{j=1}^d\partial_{y_j} [ \kappa_{ij}(\ve[y])  P_{\mathrm{s}}(\ve[y])].
\end{align}
Finally, all the terms  in Eq.~(\ref{ExpansionSquareDx}) can be evaluated by using Eqs.~(\ref{ExpressionCorruu},\ref{958I2},\ref{049141}). Taking the large time limit, we obtain the Kubo formula for the effective diffusion tensor
\begin{align}
	D_{ii}&=\int_{\Omega}d\ve[y] \  P_{\mathrm{s}}(\ve[y])\kappa_{ii}(\ve[y])\ +\nonumber\\
&\iint_{\Omega}d\ve[x]d\ve[y] \ u_i(\ve[x])G(\ve[x]\vert \ve[y])[2 J_{\mathrm{s},i}(\ve[y])-u_i(\ve[y])P_{\mathrm{s}}(\ve[y])] ,\label{ResultDiffCoeff}
\end{align}
where   $G(\ve[x]\vert \ve[y])=\int_0^{\infty}dt \{P(\ve[x],t\vert \ve[y])-P_s(\ve[x])\}$ is the pseudo-Green function \cite{barton1989elements} of  $\mathcal{L}$ on  $\Omega$.
The equation (\ref{ResultDiffCoeff}) gives in an explicit way the dispersion properties in terms of quantities that are defined at the level of an individual cell $\Omega$, with periodic boundary conditions. We may re-express $D_{ii}$  by introducing  $\ve[f](\ve[x])$, the solution of 
\begin{align}
\mathcal{L}_{\ve[x]}f_i(\ve[x]) =&- 2 J_{\mathrm{s},i}(\ve[x])+u_i(\ve[x])P_{\mathrm{s}}(\ve[x])\nonumber\\
&+P_{\mathrm{s}}(\ve[x])\int_{\Omega}d\ve[y]\ [2 J_{\mathrm{s},i}(\ve[y])-u_i(\ve[y])P_{\mathrm{s}}(\ve[y])] \label{FunctionF},
\end{align}
again with periodic boundary conditions on $\Omega$, and with the integral condition $\int_{\Omega}d\ve[x]\ \ve[f](\ve[x])=\ve[0]$. The diffusion tensor is then given by
\begin{align}
D_{ii}=&\int_{\Omega}d\ve[x] \left\{ P_{\mathrm{s}}(\ve[x])\kappa_{ii}(\ve[x])+u_i(\ve[x])f_i(\ve[x])\right\} .\label{DispersionFunctionOfF}
\end{align}
Non-equilibrium effects are manifested in Eq.~(\ref{ResultDiffCoeff}) by the presence of the local currents of the stationary state, generalizing similar Kubo formulas derived for equilibrium problems. In the case of transport by incompressible fluid flows, $P_s(\ve[x])$ is uniform,  $\ve[J]_s$ is equal to the flow $\ve[u]$ and one recovers the equations describing dispersion in incompressible hydrodynamic flows (compare for example Eqs.~(\ref{FunctionF},\ref{DispersionFunctionOfF}) to Eqs.~(35,48) of Ref. \cite{carbonell1983dispersion}).

\textit{Periodic diffusivity with an external uniform force. }
We now focus on advection-diffusion systems described by Eq.~(\ref{eqkF}), which fall in the class of the general equation (\ref{FKPEq}) with
\begin{align}
	\kappa_{ij}(\ve[x])=\delta_{ij}\kappa(\ve[x]),\ \ 	\ve[u](\ve[x])=\kappa(\ve[x]) \beta \ve[F]+\nabla\kappa(\ve[x]). 
\end{align}
The effective dispersion tensor $D_{ii}$ can be obtained by solving numerically  the partial differential equations  (\ref{FunctionF},\ref{DispersionFunctionOfF}), leading to the results on Fig.~\ref{FIG1}, which compare very well to numerical simulations of the SDE (\ref{SDE_Ito}). 

\begin{figure}[h!]
 \includegraphics[width=8cm,clip]{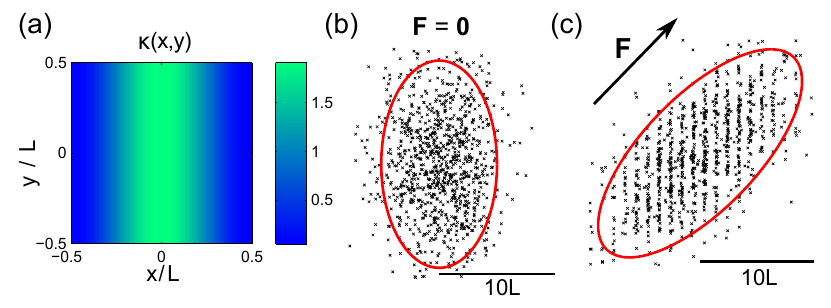}   
 \caption{(color online) (a) The 2D periodic diffusivity field for our example of stratified medium, $\kappa(x,y)=\kappa_0[1+0.95\cos(2\pi x/L)]$, shown in units of $\kappa_0$ on the fundamental rectangular unit cell. (b) and (c): Cloud of particles diffusing in the local diffusivity field shown in (a) in the presence of external force at a time $t=10L^2/\kappa_0$. In (b) no external force and in (c)  the force has magnitude  given by $\beta F L=100$, and acts in the  direction indicated by the arrow. 
The ellipses represent the region in which $95\%$ of the points should fall and are determined from Eqs. (\ref{DispersionStratesLowForce},\ref{DispersionStratesLargeForce}). 
\label{FIG2}}
\end{figure}

\textit{Stratified media. }
In systems where the local diffusivity varies only in one dimension, $\kappa(x,y)=\kappa(x)$ as illustrated in Fig.~\ref{FIG2}(a),  $\ve[f]$ depends only on $x$ and can  be calculated analytically \cite{commentStokes}. For vanishing forces, the diffusivity tensor reads
\begin{align}
	D_{xx}= 1/\overline{\kappa^{-1}}, \ D_{yy}=\overline{\kappa},\ D_{xy}=0 \hspace{0.6cm} (|\ve[F]|\rightarrow 0).\label{DispersionStratesLowForce}
\end{align}
Here the anisotropy of the dispersion is  imposed by the anisotropy of the field $\kappa$ ; from Jensen's inequality we see that    $D_{xx}\le D_{yy}$, indicating that dispersion is faster in the direction parallel to the strata of the medium [Fig. \ref{FIG2}(b)]. For  large forces however, we find that 
\begin{align}
D_{ij} = (\overline{\kappa^{-1}})^{-1}\left\{\delta_{ij}+\frac{F_i F_j}{\vert \ve[F]\cdot\ve[e]_x\vert^2}  \left[ \frac{\ \overline{\kappa^{-2}}}{(\overline{\kappa^{-1}})^{2}} -1\right]\right\},\label{DispersionStratesLargeForce}
\end{align}
so the dispersion becomes larger in the direction parallel to the force than in the perpendicular direction \cite{commentStrata}. 
The dispersion is highly sensitive to the projection of the force normal to the strata [Fig.~\ref{FIG2}(c)], and the diffusion coefficients in the planes of the strata diverge when $\ve[F]$ is in the plane of the strata (in fact they grow as $|\ve[F]|^2$). 

\textit{Force induced dispersion enhancement in 2D.} Consider the general 2D problem in the case of  large forces. For large forces, it is natural to suppose that the equilibration time in the direction (here $x$) of the force  is much shorter than in the other direction. We thus make the quasi-static approximation $P(x,y,t)\simeq \pi(y,t) P_s(x\vert y)$, where $P_s(x\vert y)\sim \kappa^{-1}(x,y)$ is the stationary probability to observe $x$ given the value of $y$. An effective Fokker-Planck equation can then be derived for the PDF $\pi(y,t)$ by integrating over $x$, and using Eqs.~(\ref{FunctionF},\ref{DispersionFunctionOfF}), to obtain  \cite{commentStokes}
\begin{align}
D_{xx}=\frac{[\beta F R(L)]^2}{W(L)}\int_0^{L} dy \left[\frac{W(y)}{W(L)}-\frac{R(y)}{R(L)}\right]^2 e^{-\overline{\ln\kappa}(y)}, \label{DxxForceSquared}
\end{align}
where $L$ is the length of the period in the direction $y$, the notation $\overline{g}(y)$ representing uniform spatial averaging over $x$ for any function $g(x,y)$, and where
\begin{align}	
	&R(y)=\int_0^{y}du\  e^{\overline{\ln\kappa}(u)}; W(y)=\int_0^{y}du\ \overline{\kappa^{-1}}(u) e^{\overline{\ln\kappa}(u)}.
\end{align}
Equation (\ref{DxxForceSquared}) shows that  local heterogeneities generically give rise to diffusion coefficients scaling as the square of the force for large forces, implying that \textit{the force-induced diffusivity can be much larger than the microscopic diffusion coefficients}. Quadrature of the integrals in Eq.~(\ref{DxxForceSquared}) give a coefficient of $F^2$ which is in agreement with the simulations, as seen in Fig.~\ref{FIG1}(d). 

\textit{Conclusion.} Taylor dispersion \cite{taylor1953dispersion} is a textbook example of a phenomenon where spatial variations of a time-independent  compressible velocity field, along with locally constant molecular diffusivity, lead to  enhanced dispersion.  Here, external uniform forces lead to increased dispersion in the direction of the force. The mechanism is similar to that behind Taylor dispersion in that particles with different trajectories experience very different advection by the applied force due to its coupling to the local mobility/diffusivity. We have also seen that an external force can non-monotonically  modify the dispersion in the direction perpendicular to the applied force. This  surprising effect is due to the fact that an applied force yields a non-uniform stationary distribution over the fundamental periodic cell. It is possible that one may construct experimental systems where the effects predicted here could be observed.  Periodic optical potentials, in which colloidal particles can be  tracked, can be generated by lasers \cite{dalle2011dynamics,Evstigneev2008} and it would be interesting to see if experimental realizations of media with spatially modulated diffusivities could be similarly produced in order to observe the effects predicted in this Letter. Finally, we stress that the results here can be applied to any periodic advection-diffusion system and thus have a wide range of applicability. For instance, one can use the formulas to study the dispersion in periodic potentials in \textit{any} dimension in the presence of an external force \cite{reimann2001giant,reimann2002diffusion} (even with varying local mobility) as well as in systems with no local detailed balance, such as active particle systems.

\newpage

\appendix
\vspace{1cm}
\begin{center}
{\bf SUPPLEMENTAL MATERIAL}
\end{center}
\section{The Generalized Stokes-Einstein Relation}
The Stokes-Einstein relation is a relationship between effective diffusivity and effective drift or mobility which applies
in equilibrium systems, and is often used to deduce diffusivity from mobility (for a recent example see \cite{putzel2014nonmonotonic}).  Here we show how, beyond the regime of linear response, the Stokes-Einstein relation breaks down due to the presence of currents associated with the stationary distribution. 

The effective drift is given by the Stratonovich formula \cite{stratonovich1958oscillator} (Eq. (6) of the main text)
\begin{equation}
V_i  =   \int_{\Omega} d\ve[x] \ P_{\mathrm{s}}(\ve[x])\  u_i(\ve[x]), 
\end{equation}
where $P_{\mathrm{s}}$ is the stationary distribution on the unit cell $\Omega$. 
Now consider a system where the local drift $u_i$ is perturbed by a small external force ${\bf F} $ so that the local drift $u_i$ changes to
\begin{equation}
u'_i({\bf x})=u_i({\bf x}) + \beta\sum_{j=1}^d\kappa_{ij}({\bf x}) F_j.
\end{equation}
The induced local drift due to the force ${\bf F}$ takes this form as the local mobility tensor is given, using the local Stokes-Einstein formula or detailed balance, by $\mu_{ij} = \beta\kappa_{ij}$, where $\beta=1/k_BT$ is the inverse of the thermal energy and $\kappa_{ij}$ is the local diffusivity tensor. From the above formulas we then see that
\begin{equation}
{\partial V_i\over \partial F_i} = \beta  \int_\Omega d{\bf x} \ P_{\mathrm{s}}(\ve[x])\kappa_{ii}({\bf x})
+ \int_\Omega d{\bf x}\  {\partial P_{\mathrm{s}}(\ve[x])\over \partial F_i}{u}_i({\bf x}).\label{intdif}
\end{equation}
Differentiating the stationary Fokker-Planck equation  $\mathcal{L}_{\ve[x]}P_{\mathrm{s}}=0$ with respect to $F_i$ then yields
\begin{equation}
\mathcal{L}_{\ve[x]}{\partial P_{\mathrm{s}}(\ve[x])\over \partial F_i} - \beta\sum_{j=1}^d{\partial \over \partial x_j}\left[\kappa_{ji}({\bf x})P_{\mathrm{s}}(\ve[x])\right]=0.\label{stokes1}
\end{equation}
The boundary conditions for $\partial P_{\mathrm{s}}(\ve[x])/\partial F_i$ are clearly that it is periodic on the boundaries of $\Omega$, but also we must have, by conservation of probability, that
\begin{equation}
\int_\Omega d{\bf x}\ {\partial P_{\mathrm{s}}(\ve[x])\over \partial F_i} = 0.\label{intc}
\end{equation} 
By definition (see \textit{e.g.} Ref. \cite{barton1989elements}), the pseudo-Green's function $G({\bf x}|{\bf y})$ for $\mathcal{L}_{\ve[x]}$ on $\Omega$ obeys
\begin{equation}
\mathcal{L}_{\ve[x]}G({\bf x}|{\bf y})= -\delta({\bf x}-{\bf y}) + P_{\mathrm{s}}(\ve[x]).
\end{equation}
We can use this  pseudo-Green's function $G({\bf x}|{\bf y})$ to construct the solution of Eq. (\ref{stokes1}) as
\begin{equation}
{\partial P_{\mathrm{s}}(\ve[x])\over \partial F_i}= -\int_\Omega d{\bf y}\ G({\bf x}|{\bf y})\beta\sum_{j=1}^d{\partial \over \partial y_j}\left[\kappa_{ji}({\bf y})P_{\mathrm{s}}(\ve[y])\right],
\end{equation}
which clearly satisfies the integral condition Eq. (\ref{intc}). Substituting this solution into Eq. (\ref{intdif}) then yields
\begin{align}
{\partial V_i\over \partial F_i} = &\beta \Big\{ \int_\Omega d{\bf x} \ P_{\mathrm{s}}(\ve[x])\kappa_{ii}({\bf x})\nonumber\\
&- \iint_\Omega d{\bf x}d{\bf y}\  {u}_i({\bf x})  G({\bf x}|{\bf y})\sum_{j=1}^d{\partial \over \partial y_j}\left[\kappa_{ji}({\bf y})P_{\mathrm{s}}(\ve[y])\right]
\Big\}.
\end{align}
Now, using Eq. (17) of the main text and the definition Eq. (16) of the current $J_{\mathrm{s},i}$ we can write
\begin{equation}
{\partial V_i\over \partial F_i} = \beta D_{ii} -\beta \iint_\Omega d{\bf x}d{\bf y} \ {u}_i({\bf x})  G({\bf x}|{\bf y})J_{\mathrm{s},i}({\bf y}).\label{mse}
\end{equation}
The Stokes Einstein relation ${\partial V_i/ \partial F_i} = \beta D_{ii}$ between the effective drift and diffusivity thus in general holds only when the current of the stationary state ${\bf J}_{\mathrm{s}}$ vanishes. We also note that the first term of Eq. (\ref{mse}), being the diffusion constant, it clearly positive. However, the sign of the second term is not obvious. 
In Ref. \cite{eichhorn2002brownian}, it was found that a particle with constant applied force in a two dimensional periodic, ratchet-like,  potential can exhibit absolute negative mobility -  it would be interesting to see if the formalism  developed here could be used to better understand this phenomenon.

\section{Results in one dimension}

In more than one  dimension the resolution of the partial differential equation Eq. (18) to evaluate the Kubo formula
for the diffusion equation is not possible analytically. However, in one dimension the corresponding differential equation can 
be evaluated analytically and we give the general result for any diffusion advection in one dimension and then specialize the result to the case of diffusion in a medium of varying diffusivity subject to an external field. Refs  \cite{lifson1962self,reimann2002diffusion,reimann2001giant,reguera2006entropic,burada2008entropic,
lindner2002,lindner2001optimal} are landmark papers in the study of diffusion in non-equilibrium systems, as the problem of diffusion in  a periodic potential plus a constant  force was first studied in Refs. \cite{reimann2002diffusion,reimann2001giant} and the result generalized to arbitrary advection diffusion was obtained in Refs. \cite{lindner2002,reguera2006entropic,burada2008entropic} (although in Ref. \cite{lindner2002} a Stratonovich prescription for the Langevin equation was used and in Ref. \cite{reguera2006entropic,burada2008entropic} a specific problem related to Fick-Jacobs diffusion was studied, the results given are in fact the most general possible in one dimension). The approach of Refs. \cite{lifson1962self,reimann2002diffusion,reimann2001giant,reguera2006entropic,burada2008entropic,
lindner2002,lindner2001optimal} was based on an expression for the diffusion constant in one dimension deduced from moments of first passage times, we will show here how the general result can be rederived via the Kubo formula Eq. (17). 

In what follows, we compute the dispersion properties for the model described by Eq. (4) of the main text in one dimension, and we show how to use these formulas to derive the effective diffusion tensor in stratified media. We use notation based on the aforementioned references to aid the reader who wished to compare the results. In one dimension, the stationary probability distribution is given by
\begin{equation}
P_{\mathrm{s}}(x)= J_{\mathrm{s}} I_+(x), 
\end{equation}
where $J_{\mathrm{s}}$ is the (constant) current in one dimension and 
\begin{align}
&I_+(x) = {\exp\left(\Gamma(x)\right)\over \kappa(x)}\int_x^\infty dx'\ \exp\left(-\Gamma(x')\right),\label{I+} \\
&\Gamma(x) = \int_0^x dx' {u(x')\over \kappa(x')}.
\end{align}
Due to the periodicity of $u$ and $\kappa$ the function $\Gamma$ obeys the relation
\begin{equation}
\Gamma(x+L) = \Gamma(x) + \Gamma(L).
\end{equation}
When $\Gamma(L)=0$  the system clearly has a steady state equilibrium distribution with no current. In writing Eq. (\ref{I+}) we have  assumed, without loss of generality, that $\Gamma(L)>0$ so that the integral on the right hand side converges. The steady state current is then obtained from the condition of normalization of $P_{\mathrm{s}}$ and is thus given by
\begin{equation}
J_ {\mathrm{s}}= {1\over \int_0^L dx I_+(x)}.\label{jo1d}
\end{equation} 
and thus the effective drift is given by $V =J_0L$.  The Eq. (18) of the main text can be solved in terms of the function $I_+$ and the function $I_-$ defined as
\begin{equation}
I_-(x)= {\exp\left(-\Gamma(x)\right)}\int_{-\infty}^x dx'\ {\exp\left(\Gamma(x')\right)\over \kappa(x')}.\label{I-}
\end{equation}
After some algebra we obtain the general compact expression for the effective large scale diffusivity
\begin{equation}
D = {L^2\int_0^L dx\ \kappa(x) I_\pm(x)^2 I_\mp (x)  \over \int_0^L dx\  I_\pm(x)^3}, \label{d1dfinal}
\end{equation} 
where $\pm$ indicates that one may (consistently) take the sign $+$ or $-$ in the above. The formula Eq. (\ref{d1dfinal}) agrees with those given in Refs. \cite{lifson1962self,reimann2002diffusion,reimann2001giant,reguera2006entropic,burada2008entropic,
lindner2002,lindner2001optimal}.

In the case of the diffusion in a periodic diffusivity field with constant applied force we find
\begin{align}
&	I_+(x) = \exp(\beta F x)\int_x^\infty dx' {\exp(-\beta F x')\over \kappa(x')},\\
	&I_-(x)= {1\over \beta F \kappa(x)}.
\end{align}
Now, we write  the inverse of $\kappa(x)$ as a Fourier series, {\em i.e.}
\begin{equation}
\kappa(x) ={1\over \overline{ \kappa^{-1}}\sum_k a_k \exp({2\pi k ix\over L}) }
\end{equation}
where $a_0=1$ and $a_{-k} =\overline{a}_k$. This then gives
\begin{equation}
I_+(x) = \overline{ \kappa^{-1}}\sum_k {a_k \exp({2\pi k ix\over L})\over \beta F-{2\pi k i\over L}},
\end{equation}
which yields the following expression for the effective diffusivity
\begin{equation}
D(F)={1\over \overline{ \kappa^{-1}}}\left[1 +2\beta^2 F^2\sum_{k>0} {|a_k|^2\over \beta^2F^2+{4\pi^2 k^2 \over L^2}}\right].
\label{onedfield}
\end{equation} 
When $F=0$ we recover  the classic result (Eq. (21)) $D(0) = \overline{ \kappa^{-1}}^{-1}$, that is to say that $D(0)$ only depends on the mean value of the inverse diffusivity. We see that for finite $F$ the diffusivity depends on all the Fourier coefficients of the inverse diffusivity, this means that in principle that measurements of the effective diffusion constant with applied external  forces could be used to reconstruct the diffusivity field in one dimension. For large $F$, $D(F)$ saturates at the value 
\begin{equation}
D(\infty)={1\over \overline{ \kappa^{-1}}}\left[1 +2\sum_{k>0} {|a_k|^2}\right] = {\overline{\kappa^{-2}}\over \overline{ \kappa^{-1}}^3}.
\end{equation}
The above formula recovers Eq. (22) for $D_{xx}$ when the force is directed in the $x$ direction (here the diffusion in the $y$ direction has no effect on that in the $x$). Note that this saturation is specific to the case of one dimension or for diffusion in stratified media  in the direction parallel to the force when there are no variations of the diffusivity in the direction perpendicular to the applied force.


\section{Details on numerical calculations and simulations.}

\textbf{Numerical solution of Eqs (18,19) of the main text. }
The numerical resolution of Eq. (18) of the main text was carried out with the finite element software FlexPDE (www.pdesolutions.com). In the examples considered the unit cell $\Omega$ was a square, with periodic boundary conditions. First the equation for the steady state distribution $P_{\mathrm{s}}(\ve[x])$  was solved, either directly or by relaxing an initially uniform probability solution to its steady state fixed point by numerically integrating the time dependent Fokker-Planck equation (in cases where there were convergence problems with the direct solution). This solution was then used to solve Eq. (18) for the two components of ${\bf f}$. Finally the diffusion coefficients $D_{xx}$ and $D_{yy}$ were obtained by numerical evaluation of the two integrals in Eq. (19) within the same software.

\textbf{Numerical simulations. } Numerical simulation of the stochastic differential equation for particles in a medium of varying diffusivity with applied external force were based upon integrating the simple discrete version of the Ito stochastic differential equation
\begin{align}
X_i(t+\Delta t) = &X_i(t) + \left[ \partial_ {x_i} \kappa({\bf X}(t)) + \beta F_i \kappa({\bf X}(t))\right] \Delta t\nonumber\\
&+ \sigma_i \sqrt{2 \kappa({\bf X}(t))\Delta t}.
\end{align}
Here  $\sigma_i$ are independent Gaussian random variables of zero mean and unit variance. Performing several runs enables the measurement of  $X_i(t)-X_i(0)$ and therefore to evaluate the effective drift and diffusivities defined in Eq. (2,3) of the main text. 
For the simulations shown in Fig.1. and Fig.2. of the main text, the time step was chosen to be $\Delta t=10^{-5}L^2/\kappa_0$, where $L$ is the size of the square unit cell and $\kappa_0$ is the diffusivity averaged over the unit cell. The effective diffusivity was obtained by evaluating the variance of $[X_i(t)-X_i(0)]/\sqrt{2t}$ and a time $t$ large enough to be in the diffusive regime (we took $t=10L^2/\kappa_0$). Averages were taken over more than $150,000$ runs, and controls were made to ensure that the simulation results do not depend on the time step. 

\section{Effective diffusivity at large forces in 2D}

In this section, we derive Eq. (23) of the main text, giving the effective diffusivity of particles submitted to a large force in a varying periodic two-dimensional diffusivity field. Here, we assume that the force is oriented in the direction $x$, and we call $h=\beta F_x$ the external field, and for simplicity, we assume that the fundamental unit cell of the structure is a rectangle of sides $L_x,L_y$.
The Fokker-Planck equation is Eq. (1) of the main text: 
\begin{align}
	\partial_t p(x,y,t)= \partial_x [\kappa(x,y) \partial_x p-  h\ \kappa(x,y)p] + \partial_y \kappa(x,y) \partial_y p. \label{FKPEq90481}
\end{align}
At high fields, $h\rightarrow\infty$, the stationary distribution $P_{\mathrm{s}}$ satisfies
\begin{align}
	0= \partial_x \left[  h\ \kappa(x,y)\ \ P_{\mathrm{s}}(x,y)\ \right]. 
\end{align}
so that the leading order term in $h$ vanishes. Therefore, the stationary distribution $P_{\mathrm{s}}$ takes the following general form:
\begin{align}
	P_{\mathrm{s}}(x,y)\simeq C(y)\kappa^{-1}(x,y), \label{9581441}
\end{align}
where $C(y)$ is a still unknown function of $y$. At high forces, it is natural to assume that the equilibration time in the direction $x$ is much shorter than the one in the direction $y$. Therefore, we approximate the propagator of the process by
\begin{align}
p(x,y,t)\simeq \pi(y,t) P_{\mathrm{s}}(x\vert y),  \label{QuasiStaticApprox}
\end{align}
where $P_{\mathrm{s}}(x\vert y)$ is the probability to observe a particle with an 
$x$-coordinate of value $x$, given that the coordinate in the other direction is $y$, and $\pi(y,t)$ is the marginal distribution of particles in the direction $y$ at time $t$. From  (\ref{9581441}) and the normalization condition, we find that
\begin{align}
	P_{\mathrm{s}}(x\vert y)=\frac{1}{\kappa(x,y)\ L_x \ \overline{\kappa^{-1}}(y)}.\label{ExprePstatXGivenY}
\end{align}
where we call $\overline{g}(y)=L_x^{-1}\int_0^{L_x} dx \ g(x,y)$ for any function $g$, with $L_x$ the length of the period in the direction $x$. Inserting the approximation (\ref{QuasiStaticApprox}) into  (\ref{FKPEq90481}) and integrating over $x$ leads to an effective Fokker-Planck equation for $\pi(y,t)$:
\begin{align}
	\partial_t \pi(y,t)\simeq \int_0^{L_x} dx  \ \partial_y \{\kappa(x,y) \partial_y [\pi(y,t)P_{\mathrm{s}}(x\vert y)]\}.
\end{align}
Performing explicitly the integral over $x$ by using (\ref{ExprePstatXGivenY}), we find
\begin{align}
\partial_t\pi(y,t)=\partial_y^2 [\kappa_e(y)\pi(y,t)]- \partial_y\{[\partial_y\overline{\ln\kappa}(y)] \kappa_e(y)\pi(y,t)\}, \label{95215}
\end{align}
where we have posed $\kappa_e(y)=1/ \overline{\kappa^{-1}}(y)$. 
For large times, the stationary distribution of the  effective Fokker-Planck equation (\ref{95215}) is
\begin{align}
\pi_{\mathrm{s}}(y)=\frac{e^{\overline{\ln\kappa}(y)}}{\kappa_e(y)\int_0^{L_y} du  \ e^{\overline{\ln\kappa}(u)}/\kappa_e(u)}.
\label{PstatEffeciveDynalmics}
\end{align} 
Now, selecting the term of order $h^2$ in the equation (19) of the main text, we get:
\begin{align}
D_{xx}\simeq &h^2 \int_0^{L_y} dy\int_0^{L_y} dy_0 \ \kappa_e(y)\nonumber\\
&\times \kappa_e(y_0)\int_0^{\infty}dt\ [\pi(y,t\vert y_0)-\pi_{\mathrm{s}}(y)] \pi_{\mathrm{s}}(y_0),\label{757815}
\end{align}
where $\pi(y,t\vert y_0)$ is the propagator for the effective dynamics in the direction $y$. It is useful to introduce the function $f_e$ defined as the solution of 
\begin{align}
	\partial_y^2 [\kappa_e(y)f_e(y)]- \partial_y\{[\partial_y\overline{\ln\kappa}(y)] \kappa_e(y)f_e(y)\}=\nonumber\\
-\kappa_e(y)\pi_{\mathrm{s}}(y)+\pi_{\mathrm{s}}(y) \int_0^{L_y} du\  \kappa_e(u)\pi_{\mathrm{s}}(u), \label{Equation_fe}
\end{align}
with the orthogonality condition $\int_0^{L_y}dy f_e(y)=0$. From Eqs. (\ref{95215},\ref{757815}), we deduce that the effective diffusion coefficient can be written in terms of $f_e$ as
\begin{align}
	D_{xx}\simeq h^2 \int_0^{L_y} dy \ \kappa_e(y) f_e(y).
\end{align}
Now, we introduce the functions $R$ and $W$ introduced in Eq. (24) of the main text: 
\begin{align}	
	&R(y)=\int_0^{y}du\  e^{\overline{\ln\kappa}(u)}; W(y)=\int_0^{y}du\ \kappa_e^{-1}(u) e^{\overline{\ln\kappa}(u)}.
\end{align}
Rewriting the right-hand side of Eq. (\ref{Equation_fe}) by using (\ref{PstatEffeciveDynalmics}) and expressing the result in terms of these two functions $R$ and $W$, we find: 
\begin{align}
	\partial_y^2 [\kappa_e(y)f_e(y)]- \partial_y\{[\partial_y\overline{\ln\kappa}(y)] \kappa_e(y)f_e(y)\}=\nonumber\\
- \frac{\partial_y R(y)}{W(L_y)}+\frac{\partial_y W(y) R(L_y)}{W(L_y)^2}.
\end{align}
Then, we can integrate once with respect to $y$. The resulting equation is a first order differential equation of a single variable function, and can be solved analytically. Taking into account the orthogonality condition $\int_0^1 dy f_e(y)=0$, we arrive after some lines of algebra, at the expression (23) of the main text. 


\begin{thebibliography}{39}
\bibitem{commentkubo}{As we are in a steady state time translation invariance holds and the earlier time $t'$ can be shifted to $0$ while the later time $t$ is shifted to $\tau = t-t'$. }
\end{thebibliography}

\end{document}